\documentclass[a4paper,11pt]{article}

\usepackage{hyperref}%
\hypersetup{%
  colorlinks=true,%
  linkcolor=blue,%
  citecolor=magenta,%
  urlcolor=black,%
  bookmarksnumbered=true,%
  bookmarkstype=toc,%
  bookmarksopen=false,%
  pdftitle={Topologically Nontrivial Three-Body Contact Interaction in One Dimension},%
  pdfkeywords={three-body problem; one dimension; topology; gauge theory},%
  pdfauthor={Satoshi Ohya}}%

\usepackage[margin=1in]{geometry}%

\usepackage[tt=false,sb]{libertine}%
\usepackage[T1]{fontenc}%
\usepackage{libertinust1math}%
\usepackage[cal=stix,scr=boondoxo,bb=boondox]{mathalpha}%
\usepackage{bm}%
\usepackage{mathtools}%
\DeclareMathOperator{\e}{e}%
\allowdisplaybreaks[1]%

\usepackage{amsthm}
\usepackage[many]{tcolorbox}
\newtheorem*{definition*}{Definition}
\tcolorboxenvironment{definition*}{
  colback=black!10!white,
  boxrule=0pt,
  boxsep=0pt,
  left=1pt,
  right=1pt,
  top=1pt,
  bottom=1pt,
  oversize=0pt,
  sharp corners,
}
\newtheorem*{proposition*}{Proposition}
\tcolorboxenvironment{proposition*}{
  colback=black!10!white,
  boxrule=0pt,
  boxsep=0pt,
  left=1pt,
  right=1pt,
  top=1pt,
  bottom=1pt,
  oversize=0pt,
  sharp corners,
}

\usepackage{tikz}

\usepackage[font=small,hang]{caption}%
\usepackage[labelformat=simple]{subcaption}%
\usepackage[backend=biber,style=phys,biblabel=brackets,block=ragged,eprint=true,url=true]{biblatex}%
\title{\Large\bfseries Topologically Nontrivial Three-Body Contact Interaction\\ in One Dimension}%
\author{\normalsize Satoshi Ohya\\[1em]
  \small\itshape Institute of Quantum Science, Nihon University,\\
  \small\itshape Kanda-Surugadai 1-8-14, Chiyoda, Tokyo 101-8308, Japan\\[1ex]
  \small\ttfamily ohya.satoshi@nihon-u.ac.jp}%
\date{\small(Dated: \today)}%

\begin{document}
\maketitle%
\flushbottom%

\begin{abstract}
  It is known that three-body contact interactions in one-dimensional
  $n(\geq3)$-body problems of nonidentical particles can be
  topologically nontrivial: they are all classified by unitary
  irreducible representations of the pure twin group $PT_{n}$. It was,
  however, unknown how such interactions are described in the
  Hamiltonian formalism. In this paper, we study topologically
  nontrivial three-body contact interactions from the viewpoint of the
  path integral. Focusing on spinless particles, we construct an
  $n(n-1)(n-2)/3!$-parameter family of $n$-body Hamiltonians that
  corresponds to one particular one-dimensional unitary representation
  of $PT_{n}$. These Hamiltonians are written in terms of background
  Abelian gauge fields that describe infinitely-thin magnetic fluxes
  in the $n$-body configuration space.
\end{abstract}

\newpage
\section{Introduction}
\label{section:1}
One of the most fundamental classes of interactions in nature would be
contact interaction: in the low-energy regime where particles'
wavelengths are longer than a characteristic range of interactions,
particles cannot resolve their microscopic details so that any
short-range interaction is expected to be described by a contact
(i.e., zero-range or pointlike) interaction. For interparticle
interactions in quantum many-body problems, there are two main
approaches to study such contact interactions.

The first approach is to consider a potential $V$ whose support is the
set of particle coincidence points $\Delta$. In this approach, contact
interactions are simply described by $V$. The second approach, on the
other hand, is to remove $\Delta$ from the many-body configuration
space $X$ and to consider the subtracted space $X-\Delta$ as a new
configuration space, because the coincidence points generally become
singularities---such as branch points---at which wavefunctions are not
well-defined. In this approach, contact interactions are described by
boundary (or connection) conditions around $\Delta$. These two
approaches basically yield the same results. However, there exists a
case in which the latter has a big advantage over the former: it is
the case where the fundamental group $\pi_{1}(X-\Delta)$ becomes
nontrivial and topology comes into play. In local quantum theory of
nonidentical particles, there are just two examples of such
topologically nontrivial contact interactions
\cite{Harshman:2018ojn,Harshman:2018wzv,Harshman:2021jlv}.\footnote{In
  one dimension, there exist topologically nontrivial but nonlocal
  $(2+2)$-body contact interactions
  \cite{Harshman:2018wzv,Harshman:2021jlv}.}

The first example is two-body contact interaction in two dimensions,
where the set of two-body coincidence points $\Delta$ becomes
codimension-two singularities in the $n$-body configuration space. In
this case, the fundamental group is given by the \textit{pure braid
  group} $PB_{n}$ \cite{Jo:1996}---a group of pure braids on $n$
strands with no double-crossings---and all the topologically
nontrivial two-body contact interactions are classified by unitary
irreducible representations of $PB_{n}$. Just as in Wilczek's
charge-flux picture of anyons \cite{Wilczek:1981du,Wilczek:1982wy},
these singularities can be viewed as infinitely-thin magnetic fluxes
and described by background gauge fields, which enter into the theory
through covariant derivatives rather than through a potential $V$.

The second example is three-body contact interaction in one dimension,
where the set of three-body coincidence points $\Delta$ also becomes
codimension-two singularities in the $n$-body configuration space. In
this case, the fundamental group is given by the \textit{pure twin
  group} $PT_{n}$ \cite{Khovanov:1996}---a group of planar pure braids
on $n$ strands with no triple-crossings---whose unitary irreducible
representations classify three-body contact interactions for
nonidentical particles in one dimension. Topological and
group-theoretical aspects of such interactions have recently been
studied by Harshman and Knapp
\cite{Harshman:2018ojn,Harshman:2018wzv,Harshman:2021jlv}. However,
the operator realization of such interactions is still missing.

The purpose of this paper is to identify the Hamiltonian operator for
topologically nontrivial three-body contact interactions in one
dimension. Focusing on spinless particles, we will show that
topologically nontrivial three-body contact interactions are
associated with background Abelian gauge fields in the configuration
space. Just as in two-body contact interactions in two dimensions,
these gauge fields describe infinitely-thin magnetic fluxes
penetrating through the codimension-two singularities in the
configuration space.

The paper is organized as follows. Section \ref{section:2} is devoted
to a detailed study of three-body problems of nonidentical spinless
particles in one dimension. Since $PT_{3}$ is isomorphic to the
additive integer group $\mathbb{Z}$
\cite{Bardakov:2019,Mostovoy:2019}, all the three-body contact
interactions are just classified by unitary representations of
$\mathbb{Z}$. We will show that, by using the path-integral formalism,
there exist two equivalent descriptions of topologically nontrivial
three-body contact interactions: the one is to impose a twisted
boundary condition around the three-body coincidence point and the
other is to introduce a background gauge field that describes an
infinitely-thin magnetic flux at the three-body coincidence point. We
will see that these two descriptions are related through a gauge
transformation and hence physically equivalent. Section
\ref{section:3} studies a generalization to $n(\geq3)$-body
problems. By generalizing the gauge-field description, we will
introduce an $n(n-1)(n-2)/3!$-parameter family of $n$-body
Hamiltonians that corresponds to one particular one-dimensional
unitary representations of $PT_{n}$. We conclude in section
\ref{section:4}. Appendix \ref{appendix:A} presents a brief review of
the pure twin group $PT_{n}$.

\section{Gauge-field description of three-body contact interaction}
\label{section:2}
In general, topologically nontrivial contact interactions originate
from nontrivial topology of many-body configuration space and are
dictated solely in terms of representation theory of the fundamental
group. This interplay between topology and representation theory is
best described by Feynman's path integral. In this section, we
illustrate this idea by focusing on three-body problems of
nonidentical spinless particles on the line $\mathbb{R}$ with
three-body contact interaction.

\subsection{Coordinate systems for the three-body configuration space}
\label{section:2.1}
To begin with, let us define some notation and coordinate systems. Let
$x_{i}\in\mathbb{R}$ be the coordinate of the $i$th particle of mass
$m_{i}$, where $i=1,2,3$. The set of three-body coincidence points is
given by the following codimension-two locus in $\mathbb{R}^{3}$:
\begin{align}
  \Delta_{3}\coloneq\{(x_{1},x_{2},x_{3})\in\mathbb{R}^{3}:x_{1}=x_{2}=x_{3}\}.\label{eq:1}
\end{align}
The three-body configuration space of nonidentical particles in one
dimension is then defined by the following subtracted space:
\begin{align}
  M_{\text{3-body}}\coloneq\mathbb{R}^{3}-\Delta_{3}.\label{eq:2}
\end{align}
In the next subsection, we will see that $\Delta_{3}$ corresponds to
the support of a fictitious infinitely-thin magnetic flux and becomes
a branch-point singularity of three-body wavefunctions.

Now, suppose that the three-body system is invariant under the spatial
translation $(x_{1},x_{2},x_{3})\mapsto(x_{1}+a,x_{2}+a,x_{3}+a)$,
where $a$ is an arbitrary real. In this case, the total momentum is
conserved so that we can separate the center-of-mass motion from the
system. The most convenient coordinate system for such a system is the
Jacobi coordinate system $(\xi_{1},\xi_{2},\xi_{3})$ given by
\begin{subequations}
  \begin{align}
    \xi_{1}&\coloneq x_{1}-x_{2},\label{eq:3a}\\
    \xi_{2}&\coloneq\frac{m_{1}(x_{1}-x_{3})+m_{2}(x_{2}-x_{3})}{m_{1}+m_{2}},\label{eq:3b}\\
    \xi_{3}&\coloneq\frac{m_{1}x_{1}+m_{2}x_{2}+m_{3}x_{3}}{m_{1}+m_{2}+m_{3}},\label{eq:3c}
  \end{align}
\end{subequations}
which satisfy the following identities:
\begin{subequations}
  \begin{align}
    \sum_{i=1}^{3}m_{i}x_{i}^{2}&=\sum_{i=1}^{3}\mu_{i}\xi_{i}^{2},\label{eq:4a}\\
    \sum_{i=1}^{3}\frac{1}{m_{i}}\frac{\partial^{2}}{\partial x_{i}^{2}}&=\sum_{i=1}^{3}\frac{1}{\mu_{i}}\frac{\partial^{2}}{\partial\xi_{i}^{2}},\label{eq:4b}
  \end{align}
\end{subequations}
where
\begin{subequations}
  \begin{align}
    \mu_{1}&\coloneq\left(\frac{1}{m_{1}}+\frac{1}{m_{2}}\right)^{-1},\label{eq:5a}\\
    \mu_{2}&\coloneq\left(\frac{1}{m_{1}+m_{2}}+\frac{1}{m_{3}}\right)^{-1},\label{eq:5b}\\
    \mu_{3}&\coloneq m_{1}+m_{2}+m_{3}.\label{eq:5c}
  \end{align}
\end{subequations}
Physically, $\xi_{1}$ is the relative coordinate between the first and
second particles; $\xi_{2}$ is the relative coordinate between the
center-of-mass of the first and second particles and the third
particle; and $\xi_{3}$ describes the center-of-mass of the three
particles. $\mu_{i}$ ($i=1,2$) is the reduced mass with respect to
$\xi_{i}$ and $\mu_{3}$ is the total mass. Noting that the condition
$x_{1}=x_{2}=x_{3}$ is equivalent to the condition
$\xi_{1}=\xi_{2}=0$, we see that the three-body configuration space
\eqref{eq:2} is factorized as follows:
\begin{align}
  M_{\text{3-body}}\cong\mathbb{R}\times\mathring{\mathbb{R}}^{2},\label{eq:6}
\end{align}
where $\mathbb{R}=\{\xi_{3}:-\infty<\xi_{3}<\infty\}$ is the
one-dimensional space of center-of-mass motion and
$\mathring{\mathbb{R}}^{2}=\{(\xi_{1},\xi_{2})\in\mathbb{R}^{2}:(\xi_{1},\xi_{2})\neq(0,0)\}$
is the one-punctured plane of relative motion. It is now obvious that
the three-body configuration space is a multiply-connected space
because there is a hole in the space of relative motion.

For the following discussion, it is convenient to introduce a polar
coordinate system $(r,\theta)$ in the one-punctured plane
$\mathring{\mathbb{R}}^{2}$. We write
\begin{subequations}
  \begin{align}
    \xi_{1}&=\sqrt{\frac{\mu_{0}}{\mu_{1}}}r\cos\theta,\label{eq:7a}\\
    \xi_{2}&=\sqrt{\frac{\mu_{0}}{\mu_{2}}}r\sin\theta,\label{eq:7b}
  \end{align}
\end{subequations}
where
\begin{subequations}
  \begin{align}
    r
    &\coloneq\sqrt{\frac{\mu_{1}\xi_{1}^{2}+\mu_{2}\xi_{2}^{2}}{\mu_{0}}}\nonumber\\
    &\,=\sqrt{\frac{\mu_{1}\xi_{1}^{2}+\mu_{2}\xi_{2}^{2}+\mu_{3}\xi_{3}^{2}-\mu_{3}\xi_{3}^{2}}{\mu_{0}}}\nonumber\\
    &\,=\sqrt{\frac{m_{1}x_{1}^{2}+m_{2}x_{2}^{2}+m_{3}x_{3}^{2}-\frac{(m_{1}x_{1}+m_{2}x_{2}+m_{3}x_{3})^{2}}{m_{1}+m_{2}+m_{3}}}{\mu_{0}}}\nonumber\\
    &\,=\sqrt{\frac{m_{1}m_{2}(x_{1}-x_{2})^{2}+m_{1}m_{3}(x_{1}-x_{3})^{2}+m_{2}m_{3}(x_{2}-x_{3})^{2}}{\mu_{0}(m_{1}+m_{2}+m_{3})}},\label{eq:8a}\\
    \theta
    &\coloneq\arctan\left(\sqrt{\frac{\mu_{2}}{\mu_{1}}}\frac{\xi_{2}}{\xi_{1}}\right)\nonumber\\
    &\,=\arctan\left(\sqrt{\frac{m_{1}m_{2}m_{3}}{m_{1}+m_{2}+m_{3}}}\frac{m_{1}(x_{1}-x_{3})+m_{2}(x_{2}-x_{3})}{m_{1}m_{2}(x_{1}-x_{2})}\right).\label{eq:8b}
  \end{align}
\end{subequations}
Here $\mu_{0}(>0)$ is an arbitrary reference mass scale introduced to
assign the dimension of length to the radius $r$. Now it is obvious
that the three-body configuration space \eqref{eq:6} can be decomposed
as
\begin{align}
  M_{\text{3-body}}\cong\mathbb{R}\times\mathbb{R}_{+}\times S^{1},\label{eq:9}
\end{align}
where $\mathbb{R}_{+}=\{r:0<r<\infty\}$ and
$S^{1}=\{\theta:0\leq\theta<2\pi\pmod{2\pi}\}$. Physically, $r$
describes the distance from the three-body coincidence point; $\theta$
describes the ratio between the center-of-mass coordinate
$\xi_{2}=(m_{1}(x_{1}-x_{3})+m_{2}(x_{2}-x_{3}))/(m_{1}+m_{2})$ and
the relative coordinate $\xi_{1}=(x_{1}-x_{3})-(x_{2}-x_{3})$ of the
first and second particles measured from the third particle's position
$x_{3}$. In the next subsection, we will see that, in the presence of
topologically nontrivial three-body contact interactions, three-body
wavefunctions become multi-valued functions of $\theta$, meaning that
there arises a branch-point singularity at the three-body coincidence
point $r=0$.

\subsection{Path integral on the three-body configuration space}
\label{section:2.2}
Now we wish to understand the physical meaning of the codimension-two
singularity. This can be done by using the path-integral formalism,
which enables us to incorporate topological effects into interaction
terms of classical Lagrangians in path-integral weights. In this
subsection, we first review the general theory of path integrals on
multiply-connected spaces and apply it to the three-body problem of
nonidentical particles in one dimension.

\subsubsection{Path integral on multiply-connected spaces: General
  theory}
\label{section:2.2.1}
Path integrals on multiply-connected spaces are best described by the
Dowker's covering-space method \cite{Dowker:1972np}. In this section,
we briefly review this method by following (and generalizing)
\cite{Ohya:2021oni,Ohya:2023ccz}.

Let $X$ be a multiply-connected space of the form $X=\widetilde{X}/G$,
where $\widetilde{X}$ is the universal covering space of $X$ and
$G\cong\pi_{1}(X)\subset\operatorname{Isom}(\widetilde{X})$ is a
discrete subgroup of the isometry of $\widetilde{X}$ that is
isomorphic to the fundamental group of $X$. The time-evolution kernel
(the integral kernel of time-evolution operator) for quantum particles
on $X$ is given by
\begin{align}
  U_{t}(x,y)=\sum_{\gamma\in G}D(\gamma)\widetilde{U}_{t}(x,\gamma y),\quad\forall x,y\in X,\label{eq:10}
\end{align}
where $\gamma y$ stands for the action of $G$ on
$y\in\widetilde{X}$. Here $D:G\to U(1)$ is a one-dimensional unitary
representation of $G$ and satisfies the following conditions for any
$\gamma,\gamma^{\prime}\in G$:
\begin{subequations}
  \begin{align}
    D(\gamma)D(\gamma^{\prime})&=D(\gamma\gamma^{\prime}),\label{eq:11a}\\
    \overline{D(\gamma)}&=D(\gamma)^{-1},\label{eq:11b}
  \end{align}
\end{subequations}
where the overline ($\overline{\phantom{m}}$) stands for the complex
conjugate. $\widetilde{U}_{t}(x,y)$ is the time-evolution kernel on
the universal covering space $\widetilde{X}$; it is given by the
Feynman's path integral:
\begin{align}
  \widetilde{U}_{t}(x,y)=\int_{q(0)=y}^{q(t)=x}\!\!\!\mathscr{D}q\exp\left(\frac{i}{\hbar}\int_{0}^{t}\!\!d\tau\,L(q(\tau),\Dot{q}(\tau))\right),\quad\forall x,y\in\widetilde{X},\label{eq:12}
\end{align}
where $L$ is a Lagrangian of classical mechanics on the universal
covering space $\widetilde{X}$. Once given the kernel $U_{t}$, the
time-evolution of a wavefunction $\psi$ is given by the map
$\psi(x)\mapsto\psi_{t}(x)$, where
\begin{align}
  \psi_{t}(x)=\int_{X}\!dy\,U_{t}(x,y)\psi(y),\quad\forall x\in X.\label{eq:13}
\end{align}
Since $\widetilde{U}_{t}$ is defined on the universal covering space,
the domain of the function $U_{t}(\cdot,\cdot)$ defined by
\eqref{eq:10} can be naturally extended from $X\times X$ to
$\widetilde{X}\times\widetilde{X}$. In particular, it satisfies the
following identity:
\begin{align}
  U_{t}(\gamma x,y)=D(\gamma)U_{t}(x,y),\quad\forall\gamma\in G,\quad\forall x,y\in\widetilde{X}.\label{eq:14}
\end{align}
Correspondingly, the domain of the wavefunction \eqref{eq:13} can also
be extended from $X$ to $\widetilde{X}$ and satisfies the following
identity:
\begin{align}
  \psi_{t}(\gamma x)=D(\gamma)\psi_{t}(x),\quad\forall\gamma\in G,\quad\forall x\in\widetilde{X}.\label{eq:15}
\end{align}
This identity turns out to give boundary conditions for quantum
systems on $X$.

\subsubsection{Path-integral description of three-body contact
  interaction}
\label{section:2.2.2}
Now let us turn to the problem of three nonidentical particles in one
dimension with a three-body contact interaction. First, the
fundamental group of the three-body configuration space \eqref{eq:9}
is isomorphic to $\pi_{1}(S^{1})$, which is given by the additive
group of integers:
\begin{align}
  \pi_{1}(M_{\text{3-body}})\cong\pi_{1}(S^{1})\cong\mathbb{Z}.\label{eq:16}
\end{align}
Physically, this fundamental group describes winding numbers of
three-particle trajectories around the codimension-two singularity
$\Delta_{3}$.

\begin{figure}[t]
  \centering%
  \input{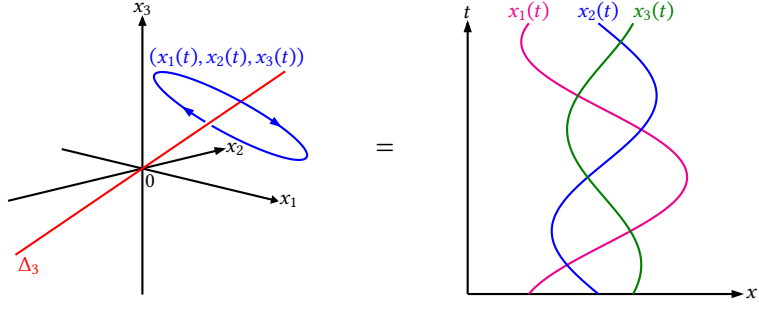}
  \caption{A three-particle trajectory with the winding number $n=1$.
    In the configuration-space picture on the left-hand side, such a
    trajectory is described by a closed loop encircling the
    codimension-two singularity $\Delta_{3}$. In the space-time
    picture on the right-hand side, such a trajectory is described by
    a planar pure braid on three strands. For planar pure braids, see
    appendix \ref{appendix:A}.}
  \label{figure:1}
\end{figure}

Next, we have to find one-dimensional unitary representations of
$\mathbb{Z}$. Since $\mathbb{Z}$ is a free group generated by a single
generator $\gamma$, there exists a one-parameter family of
one-dimensional unitary representations of $\mathbb{Z}$ labeled by a
real parameter $\alpha$. It is given by the map
$D^{[\alpha]}:\mathbb{Z}\to U(1)$, where
\begin{align}
  D^{[\alpha]}(\gamma)=\exp(i2\pi\alpha),\quad\alpha\in[0,1).\label{eq:17}
\end{align}
Correspondingly, for three nonidentical particles in one dimension,
there exists a one-parameter family of time-evolution kernels given by
\begin{align}
  U^{[\alpha]}_{t}(x,y)=\sum_{n=-\infty}^{\infty}D^{[\alpha]}(\gamma^{n})\int_{q(0)=\gamma^{n}y}^{q(t)=x}\!\!\!\mathscr{D}q\exp\left(\frac{i}{\hbar}\int_{0}^{t}\!\!d\tau\,L(q(\tau),\Dot{q}(\tau))\right),\label{eq:18}
\end{align}
which satisfies the following twisted boundary condition (see
\eqref{eq:14}):
\begin{align}
  U^{[\alpha]}_{t}(\gamma x,y)=D^{[\alpha]}(\gamma)U^{[\alpha]}_{t}(x,y).\label{eq:19}
\end{align}
Note that, in the polar coordinate system introduced in section
\ref{section:2.1}, the action of $\gamma^{n}\in\mathbb{Z}$ is written
as $(r,\theta,\xi_{3})\mapsto(r,\theta+2n\pi,\xi_{3})$. It is now
clear that the one-dimensional unitary representation $D^{[\alpha]}$
describes the Aharonov-Bohm phase acquired by a three-particle state
encircling the codimension-two singularity; see figure
\ref{figure:1}. It is also clear from \eqref{eq:15} that the
three-body wavefunction becomes a multi-valued function of $\theta$
with the branch-point singularity at $r=0$.

Now, there exists an alternative, unitary-equivalent description to
this three-body system. To see this, let us consider a gauge
transformation
$\psi(x)\mapsto\psi^{\prime}(x)=W(C_{x,x_{0}})^{-1}\psi(x)$, where
$W(C_{x,x_{0}})$ is the Wilson line given by
\begin{align}
  W(C_{x,x_{0}})=\exp\left(\frac{i}{\hbar}\int_{C_{x,x_{0}}}\!\!A\right)=\e^{i\alpha(\theta(x)-\theta(x_{0}))}.\label{eq:20}
\end{align}
Here
$C_{x,x_{0}}=\{q(\tau)\in
M_{\text{3-body}}:q(0)=x_{0},~q(t)=x,~\tau\in[0,t]\}$ is a path from
$x_{0}$ to $x$ with $x_{0}$ being an arbitrary reference
point. $\theta(x)$ stands for the angle defined by the last line of
\eqref{eq:8b}. $A$ is a gauge-field one-form given by
\begin{align}
  A
  &=\hbar\alpha d\theta\nonumber\\
  &=\hbar\alpha\frac{\sqrt{\mu_{1}\mu_{2}}\left(-\xi_{2}d\xi_{1}+\xi_{1}d\xi_{2}\right)}{\mu_{1}\xi_{1}^{2}+\mu_{2}\xi_{2}^{2}}\nonumber\\
  &=\sum_{i=1}^{3}A_{i}dx_{i},\label{eq:21}
\end{align}
where
\begin{align}
  A_{i}
  &=\hbar\alpha\frac{\partial\theta}{\partial x_{i}}\nonumber\\
  &=-\frac{\hbar\alpha\sqrt{m_{1}m_{2}m_{3}(m_{1}+m_{2}+m_{3})}(\delta_{i1}(x_{2}-x_{3})+\delta_{i2}(x_{3}-x_{1})+\delta_{i3}(x_{1}-x_{2}))}{m_{1}m_{2}(x_{1}-x_{2})^{2}+m_{2}m_{3}(x_{2}-x_{3})^{2}+m_{3}m_{1}(x_{3}-x_{1})^{2}}.\label{eq:22}
\end{align}
Under this gauge transformation, eq.~\eqref{eq:18} is transformed as
$U^{[\alpha]}_{t}(x,y)\mapsto U^{[\alpha]\prime}_{t}(x,y)$, where
\begin{align}
  U^{[\alpha]\prime}_{t}(x,y)
  &=W(C_{x,x_{0}})^{-1}U^{[\alpha]}_{t}(x,y)W(C_{y,x_{0}})\nonumber\\
  &=\e^{-i\alpha(\theta(x)-\theta(y))}U^{[\alpha]}_{t}(x,y)\nonumber\\
  &=\sum_{n=-\infty}^{\infty}\e^{-i\alpha(\theta(x)-\theta(y)-2n\pi)}\int_{q(0)=\gamma^{n}y}^{q(t)=x}\!\!\!\mathscr{D}q\exp\left(\frac{i}{\hbar}\int_{0}^{t}\!\!d\tau\,L(q(\tau),\Dot{q}(\tau))\right)\nonumber\\
  &=\sum_{n=-\infty}^{\infty}\int_{q(0)=\gamma^{n}y}^{q(t)=x}\!\!\!\mathscr{D}q\exp\left(\frac{i}{\hbar}\int_{0}^{t}\!\!d\tau\left[L(q(\tau),\Dot{q}(\tau))-\sum_{i=1}^{3}\Dot{q}_{i}(\tau)A_{i}(q(\tau))\right]\right).\label{eq:23}
\end{align}
Here the last line follows from
$\exp(-(i/\hbar)\int_{0}^{t}\!d\tau\sum_{i=1}^{3}\Dot{q}_{i}A_{i})=\exp(-(i/\hbar)\int_{C_{q(t),q(0)}}\!dA)=\e^{-i\alpha(\theta(q(t))-\theta(q(0)))}$
and
$\theta(q(t))-\theta(q(0))=\theta(x)-\theta(\gamma^{n}y)=\theta(x)-\theta(y)-2n\pi$
for a path $q(\tau)$ that satisfies the conditions $q(0)=\gamma^{n}y$
and $q(t)=x$. Notice that, in this new gauge, the time-evolution
kernel \eqref{eq:23} satisfies the periodic boundary condition
\begin{align}
  U^{[\alpha]\prime}_{t}(\gamma x,y)=U^{[\alpha]\prime}_{t}(x,y).\label{eq:24}
\end{align}
Note also that the gauge-field one-form \eqref{eq:21} describes an
infinitely-thin magnetic flux penetrating through the three-body
coincidence point $x_{1}=x_{2}=x_{3}$ (or $\xi_{1}=\xi_{2}=0$) in
$M_{\text{3-body}}$. In fact, the field-strength two-form $F=dA$ takes
the following form:
\begin{align}
  F
  &=2\pi\hbar\alpha\delta(\xi_{1})\delta(\xi_{2})d\xi_{1}\wedge d\xi_{2}\nonumber\\
  &=2\pi\hbar\alpha\delta(x_{1}-x_{2})\delta(x_{2}-x_{3})\left(dx_{1}\wedge dx_{2}+dx_{2}\wedge dx_{3}+dx_{3}\wedge dx_{1}\right),\label{eq:25}
\end{align}
which describes the infinitely-thin magnetic field along the
codimension-two singularity $\Delta_{3}$. Hence, there exist two
equivalent descriptions for topologically nontrivial three-body
contact interactions: the first is the twisted boundary condition
\eqref{eq:19} with no gauge field, and the second is the periodic
boundary condition \eqref{eq:24} with the background Abelian gauge
field \eqref{eq:22}. These two descriptions are unitarily equivalent
so that they describe the same physical system.

\subsubsection{Hamiltonian description of three-body contact
  interaction}
\label{section:2.2.3}
Let us finally study Hamiltonian descriptions of the system. Under the
twisted boundary condition \eqref{eq:19}, the classical Hamiltonian is
just given by the Legendre transform of the classical Lagrangian
$L(x,\Dot{x})$ in the path integral \eqref{eq:18}. If $L$ is of the
form
$L(x,\Dot{x})=\sum_{i=1}^{3}(m_{i}/2)\Dot{x}_{i}^{2}-V(x_{1},x_{2},x_{3})$,
then the corresponding Hamiltonian operator in the quantum three-body
problem is just given by
\begin{align}
  H_{\text{twisted}}=-\sum_{i=1}^{3}\frac{\hbar^{2}}{2m_{i}}\frac{\partial^{2}}{\partial x_{i}^{2}}+V(x_{1},x_{2},x_{3}).\label{eq:26}
\end{align}
On the other hand, under the periodic boundary condition
\eqref{eq:24}, the classical Hamiltonian is given by the Legendre
transform of the classical Lagrangian
$L(x,\Dot{x})-\sum_{i=1}^{3}\Dot{x}_{i}A_{i}=\sum_{i=1}^{3}(m_{i}/2)\Dot{x}_{i}^{2}-V(x_{1},x_{2},x_{3})-\sum_{i=1}^{3}\Dot{x}_{i}A_{i}$
in the path integral \eqref{eq:23}. The corresponding Hamiltonian
operator then becomes
\begin{align}
  H_{\text{periodic}}=-\sum_{i=1}^{3}\frac{\hbar^{2}}{2m_{i}}\left(\frac{\partial}{\partial x_{i}}+\frac{i}{\hbar}A_{i}\right)^{2}+V(x_{1},x_{2},x_{3}).\label{eq:27}
\end{align}
Note that these two operators are unitarily equivalent and mutually
related through the unitary transformation
$H_{\text{periodic}}=W(C_{x,x_{0}})^{-1}H_{\text{twisted}}W(C_{x,x_{0}})$.

\section{A generalization to \texorpdfstring{$n$}{n}-body problems}
\label{section:3}
Now we wish to generalize the results in the previous section to
$n(\geq3)$-body problems of nonidentical particles. Obviously, the key
is the fundamental group of $n$-body configuration space. Under
three-body contact interactions, the $n$-body configuration space of
nonidentical particles on $\mathbb{R}$ is given by
\begin{align}
  M_{\text{$n$-body}}\coloneq\mathbb{R}^{n}-\Delta_{n},\label{eq:28}
\end{align}
where $\Delta_{n}$ is the set of three-body coincidence points defined
by
\begin{align}
  \Delta_{n}\coloneq\{(x_{1},\cdots,x_{n})\in\mathbb{R}^{n}:x_{i}=x_{j}=x_{k}\quad(1\leq i<j<k\leq n)\}.\label{eq:29}
\end{align}
The fundamental group of this space has been studied in the
mathematical literature: in the late 1990s, Khovanov
\cite{Khovanov:1996} showed that the fundamental group of
\eqref{eq:28} is isomorphic to the pure twin group $PT_{n}$---a group
of planar pure braids on $n$ strands without triple-intersection
points. Thus we find
\begin{align}
  \pi_{1}(M_{\text{$n$-body}})\cong PT_{n}.\label{eq:30}
\end{align}
In appendix \ref{appendix:A}, we give a brief review of $PT_{n}$ and
\eqref{eq:30}. Note that $PT_{3}$ is isomorphic to $\mathbb{Z}$
\cite{Bardakov:2019,Mostovoy:2019}.

Hence, to identify topologically nontrivial contact interactions, we
have to classify all possible one-dimensional unitary representations
of $PT_{n}$. This is, however, a very complicated mathematical problem
and in fact still unsolved. In the following, we would like to
construct a family of $n$-body Hamiltonians that corresponds to one
particular, yet physically reasonable, unitary representation of
$PT_{n}$ by using physical intuition.

First, in the three-body problem, the topologically nontrivial
three-body contact interaction was described by a single
infinitely-thin magnetic flux penetrating through the three-body
coincidence point $x_{1}=x_{2}=x_{3}$. In the $n$-body problem, there
exist $\binom{n}{3}=n(n-1)(n-2)/3!$ distinct three-body coincidence
points given by $x_{i}=x_{j}=x_{k}$ with $1\leq i<j<k\leq n$. Hence,
it would be natural to expect that topologically nontrivial three-body
contact interactions in the $n$-body problem could be described by
$n(n-1)(n-2)/3!$ distinct infinitely-thin magnetic fluxes penetrating
through these three-body coincidence points. The field-strength
two-form describing this situation is easily found to be
\begin{align}
  F=2\pi\hbar\sum_{1\leq i<j<k\leq n}\alpha_{ijk}\delta(x_{i}-x_{j})\delta(x_{j}-x_{k})\left(dx_{i}\wedge dx_{j}+dx_{j}\wedge dx_{k}+dx_{k}\wedge dx_{i}\right),\label{eq:31}
\end{align}
where $\alpha_{ijk}$ is a real parameter. This two-form can be written
as $F=dA$, where
\begin{align}
  A=\hbar\sum_{1\leq i<j<k\leq n}\alpha_{ijk}d\theta_{ijk}.\label{eq:32}
\end{align}
Here $\theta_{ijk}$ is a natural generalization of \eqref{eq:8b} given
by
\begin{align}
  \theta_{ijk}\coloneq\arctan\left(\sqrt{\frac{m_{i}m_{j}m_{k}}{m_{i}+m_{j}+m_{k}}}\frac{m_{i}(x_{i}-x_{k})+m_{j}(x_{j}-x_{k})}{m_{i}m_{j}(x_{i}-x_{j})}\right).\label{eq:33}
\end{align}
Notice that, in the Cartesian coordinate system
$(x_{1},\cdots,x_{n})$, the gauge-field one-form \eqref{eq:32} is
written as $A=\sum_{i=1}^{n}A_{i}dx_{i}$, where
\begin{align}
  A_{i}
  &=\hbar\sum_{1\leq j<k<l\leq n}\alpha_{jkl}\frac{\partial\theta_{jkl}}{\partial x_{i}}\nonumber\\
  &=-\sum_{1\leq j<k<l\leq n}\frac{\hbar\alpha_{jkl}\sqrt{m_{j}m_{k}m_{l}(m_{j}+m_{k}+m_{l})}(\delta_{ij}(x_{k}-x_{l})+\delta_{ik}(x_{l}-x_{j})+\delta_{il}(x_{j}-x_{k}))}{m_{j}m_{k}(x_{j}-x_{k})^{2}+m_{k}m_{l}(x_{k}-x_{l})^{2}+m_{l}m_{j}(x_{l}-x_{j})^{2}}.\label{eq:34}
\end{align}
The $n$-body Hamiltonian operator in the presence of the fictitious
magnetic fluxes \eqref{eq:31} is then given by
\begin{align}
  H_{\text{periodic}}=-\sum_{i=1}^{n}\frac{\hbar^{2}}{2m_{i}}\left(\frac{\partial}{\partial x_{i}}+\frac{i}{\hbar}A_{i}\right)^{2}+V(x_{1},\cdots,x_{n}).\label{eq:35}
\end{align}
This is the simplest one-dimensional $n$-body model of nonidentical
particles with topologically nontrivial three-body contact
interactions. Notice that, in this gauge-field description, $n$-body
wavefunctions satisfy the periodic boundary condition around the
three-body coincidence points.

Finally, we note that the model \eqref{eq:35} corresponds to the
$n(n-1)(n-2)/3!$-parameter family of one-dimensional unitary
representations $D^{[\alpha_{ijk}]}:PT_{n}\to U(1)$ given by the
following Wilson loop:
\begin{align}
  D^{[\alpha_{ijk}]}(\gamma)=\exp\left(\frac{i}{\hbar}\oint_{\gamma}\!A\right),\quad\forall\gamma\in PT_{n}\cong\pi_{1}(M_{\text{$n$-body}}).\label{eq:36}
\end{align}
This Wilson loop clearly satisfies the group multiplication law
$D^{[\alpha_{ijk}]}(\gamma)D^{[\alpha_{ijk}]}(\gamma^{\prime})=D^{[\alpha_{ijk}]}(\gamma\gamma^{\prime})$
and the unitarity
$\overline{D^{[\alpha_{ijk}]}(\gamma)}=D^{[\alpha_{ijk}]}(\gamma)^{-1}=D^{[\alpha_{ijk}]}(\gamma^{-1})$
for any planar pure braids $\gamma,\gamma^{\prime}\in PT_{n}$ (which
are in one-to-one correspondence with closed loops
$\gamma,\gamma^{\prime}$ in $M_{\text{$n$-body}}$), thus giving a
one-dimensional unitary representation of $PT_{n}$. Note, however,
that eq.~\eqref{eq:36} is just one particular example: for instance,
$PT_{4}$ is known to be isomorphic to the free group $F_{7}$
\cite{Bardakov:2019,Mostovoy:2019} so that its one-dimensional unitary
representation contains up to seven independent parameters
\cite{Harshman:2018wzv}, while \eqref{eq:36} possesses just four
parameters for $n=4$. Our simple magnetic fluxes \eqref{eq:31} are
therefore just a small subset of all possible three-body contact
interactions. Future studies should investigate the most general
gauge-field one-form that describes the entire one-dimensional unitary
representation of $PT_{n}$.

\section{Conclusion}
\label{section:4}
As was shown by Harshman and Knapp
\cite{Harshman:2018ojn,Harshman:2018wzv,Harshman:2021jlv}, three-body
contact interactions in $n$-body problems of nonidentical particles on
$\mathbb{R}$ can be topologically nontrivial: they are all classified
by unitary irreducible representations of the pure twin group
$PT_{n}$---the fundamental group of the $n$-body configuration space
$M_{\text{$n$-body}}=\mathbb{R}^{n}-\Delta_{n}$. It was, however,
unknown how those topologically nontrivial three-body contact
interactions are described by Hamiltonian operators.

This paper studied Hamiltonian descriptions for the topologically
nontrivial three-body contact interactions by using the path-integral
formalism. In the three-body problem, we showed that all the
three-body contact interactions corresponding to one-dimensional
unitary representations of the fundamental group
$\pi_{1}(M_{\text{3-body}})\cong PT_{3}\cong\mathbb{Z}$ are realized
by background Abelian gauge fields. These gauge fields describe an
infinitely-thin magnetic flux penetrating through the three-body
coincidence point in $M_{\text{3-body}}$. By generalizing this result,
we constructed the $n(n-1)(n-2)/3!$-parameter family of
$n(\geq3)$-body Hamiltonians for nonidentical particles on
$\mathbb{R}$. This family is made up of background Abelian gauge
fields that describe $n(n-1)(n-2)/3!$ infinitely-thin magnetic fluxes
in $M_{\text{$n$-body}}$ and corresponds to one-dimensional unitary
representations of $\pi_{1}(M_{\text{$n$-body}})\cong PT_{n}$ given by
\eqref{eq:36}.

There remain several issues still to be addressed. Examples include
the classification of unitary irreducible representations of $PT_{n}$,
the construction of field-theory description for topologically
nontrivial three-body contact interactions, and experimental
realizations. We hope to address these issues in the future.

\subsection*{Acknowledgment}
This work was supported by JSPS KAKENHI Grant Number JP23K03267.

\appendix
\setcounter{equation}{0}
\renewcommand{\theequation}{\thesection.\arabic{equation}}
\section{Pure twin group}
\label{appendix:A}
The pure twin group $PT_{n}$ is defined as a subgroup of the
\textit{twin group} $T_{n}$---the group of planar braids on $n$
strands with no triple-crossings.\footnote{The (pure) twin group is
  also referred to as the planar (pure) braid group
  \cite{Mostovoy:2019} and the (pure) traid group
  \cite{Harshman:2018ojn,Harshman:2018wzv,Harshman:2021jlv}.} In this
appendix, we give a brief informal review of $T_{n}$ and
$PT_{n}$. Following the original work of Khovanov \cite{Khovanov:1996}
(with terminology borrowed from \cite{Mostovoy:2019}), we would like
to explain that $PT_{n}$ is isomorphic to the fundamental group of the
$n$-body configuration space \eqref{eq:28}.

Let us start with a diagrammatic definition of the twin group
$T_{n}$. First, a planar braid on $n$ strands is an equivalence class
of planar diagrams like the following:
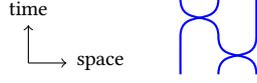
\begin{center}
  \begin{tikzpicture}[baseline=-0.65ex,scale=0.5]
    \draw[black,->] (-4,-0.7) -- (-3,-0.7) node[font=\scriptsize,right]{space};
    \draw[black,->] (-4,-0.7) -- (-4,0.3) node[font=\scriptsize,above]{time};
    \draw[blue,thick] (0,-1) .. controls +(0,0.8) and +(0,-0.8) .. (0,0) .. controls +(0,0.8) and +(0,-0.8) .. (1,1);
    \draw[blue,thick] (1,-1) .. controls +(0,0.8) and +(0,-0.8) .. (2,0) .. controls +(0,0.8) and +(0,-0.8) .. (2,1);
    \draw[blue,thick] (2,-1) .. controls +(0,0.8) and +(0,-0.8) .. (1,0) .. controls +(0,0.8) and +(0,-0.8) .. (0,1);
  \end{tikzpicture}
\end{center}
This kind of planar diagrams can be viewed as representing worldlines
of point particles in two-dimensional spacetime, where an intersection
point of $n$ worldlines represent a spacetime point of $n$-particle
collision. Since we are interested in the topological structure of
planar braids, we allow each worldline to deform continuously in the
spacetime (keeping the initial and final points fixed). In particular,
we allow two worldlines to intersect at a single point
like \begin{tikzpicture} \draw[blue,thick] (0ex,0ex) -- (2ex,1.2ex);
  \draw[blue,thick] (0ex,1.2ex) -- (2ex,0ex); \end{tikzpicture}\,. But
we do not allow three worldlines to intersect at a single point like
\begin{tikzpicture} \draw[blue,thick] (0ex,0ex) -- (2ex,1.2ex);
  \draw[blue,thick] (0ex,1.2ex) -- (2ex,0ex); \draw[blue,thick]
  (1ex,-0.1ex) -- (1ex,1.3ex); \end{tikzpicture}\,, because we want to
exclude three-body coincidence points. With this intersection rule,
planar braids on $n$ strands form a group, called the twin group
$T_{n}$, where the group multiplication is the concatenation of
strands, the identity element is an equivalence class of the
non-intersecting strands, and the inverse of a planar braid is its
time reversal; see below.
\begin{subequations}
  \begin{align}
    \text{group multiplication:}
    &\quad
      \begin{tikzpicture}[baseline=-0.65ex,scale=0.5]
        \draw[blue,thick] (0,-1) .. controls +(0,0.8) and +(0,-0.8) .. (1,1);
        \draw[blue,thick] (1,-1) .. controls +(0,0.8) and +(0,-0.8) .. (0,1);
        \draw[blue,thick] (2,-1) .. controls +(0,0.8) and +(0,-0.8) .. (2,1);
      \end{tikzpicture}
      ~~\times~~
      \begin{tikzpicture}[baseline=-0.65ex,scale=0.5]
        \draw[blue,thick] (0,-1) .. controls +(0,0.8) and +(0,-0.8) .. (0,1);
        \draw[blue,thick] (1,-1) .. controls +(0,0.8) and +(0,-0.8) .. (2,1);
        \draw[blue,thick] (2,-1) .. controls +(0,0.8) and +(0,-0.8) .. (1,1);
      \end{tikzpicture}
      ~~=~~
      \begin{tikzpicture}[baseline=-0.65ex,scale=0.5]
        \draw[blue,thick] (0,-1) .. controls +(0,0.8) and +(0,-0.8) .. (0,0) .. controls +(0,0.8) and +(0,-0.8) .. (1,1);
        \draw[blue,thick] (1,-1) .. controls +(0,0.8) and +(0,-0.8) .. (2,0) .. controls +(0,0.8) and +(0,-0.8) .. (2,1);
        \draw[blue,thick] (2,-1) .. controls +(0,0.8) and +(0,-0.8) .. (1,0) .. controls +(0,0.8) and +(0,-0.8) .. (0,1);
      \end{tikzpicture}\\
    \text{identity element:}
    &\quad
      \begin{tikzpicture}[baseline=-0.65ex,scale=0.5]
        \draw[blue,thick] (0,-1) .. controls +(0,0.8) and +(0,-0.8) .. (0,0) .. controls +(0,0.8) and +(0,-0.8) .. (0,1);
        \draw[blue,thick] (1,-1) .. controls +(0,0.8) and +(0,-0.8) .. (1,0) .. controls +(0,0.8) and +(0,-0.8) .. (1,1);
        \draw[blue,thick] (2,-1) .. controls +(0,0.8) and +(0,-0.8) .. (2,0) .. controls +(0,0.8) and +(0,-0.8) .. (2,1);
      \end{tikzpicture}\\
    \text{inverse:}
    &\quad
      \left(~~\begin{tikzpicture}[baseline=-0.65ex,scale=0.5]
        \draw[blue,thick] (0,-1) .. controls +(0,0.8) and +(0,-0.8) .. (0,0) .. controls +(0,0.8) and +(0,-0.8) .. (1,1);
        \draw[blue,thick] (1,-1) .. controls +(0,0.8) and +(0,-0.8) .. (2,0) .. controls +(0,0.8) and +(0,-0.8) .. (2,1);
        \draw[blue,thick] (2,-1) .. controls +(0,0.8) and +(0,-0.8) .. (1,0) .. controls +(0,0.8) and +(0,-0.8) .. (0,1);
      \end{tikzpicture}~~\right)^{-1}
      ~~=~~
      \begin{tikzpicture}[baseline=-0.65ex,scale=0.5]
        \draw[blue,thick] (0,-1) .. controls +(0,0.8) and +(0,-0.8) .. (1,0) .. controls +(0,0.8) and +(0,-0.8) .. (2,1);
        \draw[blue,thick] (1,-1) .. controls +(0,0.8) and +(0,-0.8) .. (0,0) .. controls +(0,0.8) and +(0,-0.8) .. (0,1);
        \draw[blue,thick] (2,-1) .. controls +(0,0.8) and +(0,-0.8) .. (2,0) .. controls +(0,0.8) and +(0,-0.8) .. (1,1);
      \end{tikzpicture}
  \end{align}
\end{subequations}
From the group multiplication, it is clear that any planar braids on
$n$ strands with $N$ intersection points can be written as a product
$t_{i_{1}}t_{i_{2}}\cdots t_{i_{N}}$, where $t_{i}$ ($i=1,\cdots,n-1$)
is an equivalence class of the following planar braid:
\begin{align}
  t_{i}~~=~~
  \begin{tikzpicture}[baseline=-0.65ex,scale=0.5]
    \draw[blue,thick] (-2,-1) node[black,font=\scriptsize,below]{$1$} .. controls +(0,0.8) and +(0,-0.8) .. (-2,1);
    \draw[blue,thick] (-1,-1) .. controls +(0,0.8) and +(0,-0.8) .. (-1,1);
    \draw[blue,thick] (0,-1) .. controls +(0,0.8) and +(0,-0.8) .. (0,1);
    \draw[blue,thick] (1,-1) node[black,font=\scriptsize,below]{$i$} .. controls +(0,0.8) and +(0,-0.8) .. (2,1);
    \draw[blue,thick] (2,-1) node[black,font=\scriptsize,below]{~~$i+1$} .. controls +(0,0.8) and +(0,-0.8) .. (1,1);
    \draw[blue,thick] (3,-1) .. controls +(0,0.8) and +(0,-0.8) .. (3,1);
    \draw[blue,thick] (4,-1) .. controls +(0,0.8) and +(0,-0.8) .. (4,1);
    \draw[blue,thick] (5,-1) node[black,font=\scriptsize,below]{$n$} .. controls +(0,0.8) and +(0,-0.8) .. (5,1);
  \end{tikzpicture}
\end{align}
These planar braids play the roles of generators of $T_{n}$.

Now we wish to know algebraic relations among the generators
$\{t_{1},\cdots,t_{n-1}\}$. Obviously, two twice-intersecting
worldlines can be continuously deformed to two non-intersecting
worldlines without making triple intersections. Thus we have the
identity $t_{i}^{2}=1$:
\begin{align}
  \begin{tikzpicture}[baseline=-0.3ex,scale=0.5]
    \draw[blue,thick] (-1,0) node[blue]{$\cdots$};
    \draw[blue,thick] (0,-1) node[black,font=\scriptsize,below]{$i$} .. controls +(0,0.8) and +(0,-0.8) .. (1,0) .. controls +(0,0.8) and +(0,-0.8) .. (0,1);
    \draw[blue,thick] (1,-1) node[black,font=\scriptsize,below]{~~$i+1$} .. controls +(0,0.8) and +(0,-0.8) .. (0,0) .. controls +(0,0.8) and +(0,-0.8) .. (1,1);
    \draw[blue,thick] (2,0) node[blue]{$\cdots$};
  \end{tikzpicture}
  ~~=~~
  \begin{tikzpicture}[baseline=-0.3ex,scale=0.5]
    \draw[blue,thick] (-1,0) node[blue]{$\cdots$};
    \draw[blue,thick] (0,-1) node[black,font=\scriptsize,below]{$i$} .. controls +(0,0.8) and +(0,-0.8) .. (0,1);
    \draw[blue,thick] (1,-1) node[black,font=\scriptsize,below]{~~$i+1$} .. controls +(0,0.8) and +(0,-0.8) .. (1,1);
    \draw[blue,thick] (2,0) node[blue]{$\cdots$};
  \end{tikzpicture}
\end{align}
It is also obvious that, if $|i-j|>2$, planar braids $t_{i}t_{j}$ and
$t_{j}t_{i}$ can be continuously deformed into one another without
making triple intersections. Thus we have the commutation relation
$t_{i}t_{j}=t_{j}t_{i}$ for $|i-j|>2$:
\begin{align}
  \begin{tikzpicture}[baseline=-0.3ex,scale=0.5]
    \draw[blue,thick] (-1,0) node[blue]{$\cdots$};
    \draw[blue,thick] (0,-1) node[black,font=\scriptsize,below]{$i$} .. controls +(0,0.8) and +(0,-0.8) .. (0,0) .. controls +(0,0.8) and +(0,-0.8) .. (1,1);
    \draw[blue,thick] (1,-1) node[black,font=\scriptsize,below]{$i+1$} .. controls +(0,0.8) and +(0,-0.8) .. (1,0) .. controls +(0,0.8) and +(0,-0.8) .. (0,1);
    \draw[blue,thick] (2,0) node[blue]{$\cdots$};
    \draw[blue,thick] (3,-1) node[black,font=\scriptsize,below]{$j$} .. controls +(0,0.8) and +(0,-0.8) .. (4,0) .. controls +(0,0.8) and +(0,-0.8) .. (4,1);
    \draw[blue,thick] (4,-1) node[black,font=\scriptsize,below]{$j+1$} .. controls +(0,0.8) and +(0,-0.8) .. (3,0) .. controls +(0,0.8) and +(0,-0.8) .. (3,1);
    \draw[blue,thick] (5,0) node[blue]{$\cdots$};
  \end{tikzpicture}
  ~~=~~
  \begin{tikzpicture}[baseline=-0.3ex,scale=0.5]
    \draw[blue,thick] (-1,0) node[blue]{$\cdots$};
    \draw[blue,thick] (0,-1) node[black,font=\scriptsize,below]{$i$} .. controls +(0,0.8) and +(0,-0.8) .. (1,0) .. controls +(0,0.8) and +(0,-0.8) .. (1,1);
    \draw[blue,thick] (1,-1) node[black,font=\scriptsize,below]{$i+1$} .. controls +(0,0.8) and +(0,-0.8) .. (0,0) .. controls +(0,0.8) and +(0,-0.8) .. (0,1);
    \draw[blue,thick] (2,0) node[blue]{$\cdots$};
    \draw[blue,thick] (3,-1) node[black,font=\scriptsize,below]{$j$} .. controls +(0,0.8) and +(0,-0.8) .. (3,0) .. controls +(0,0.8) and +(0,-0.8) .. (4,1);
    \draw[blue,thick] (4,-1) node[black,font=\scriptsize,below]{$j+1$} .. controls +(0,0.8) and +(0,-0.8) .. (4,0) .. controls +(0,0.8) and +(0,-0.8) .. (3,1);
    \draw[blue,thick] (5,0) node[blue]{$\cdots$};
  \end{tikzpicture}
\end{align}
Notice that, for $|i-j|=1$, there is no such relation between
$t_{i}t_{j}$ and $t_{j}t_{i}$. In particular, $t_{i+1}t_{i}t_{i+1}$
and $t_{i}t_{i+1}t_{i}$ cannot be continuously deformed to one another
without making triple intersections; that is, the Yang-Baxter relation
does not hold in the twin group:
\begin{align}
  \begin{tikzpicture}[baseline=-0.3ex,scale=0.5]
    \draw[blue,thick] (-1,0) node[blue]{$\cdots$};
    \draw[blue,thick] (0,-1.5) node[black,font=\scriptsize,below]{$i$} .. controls +(0,0.8) and +(0,-0.8) .. (0,-0.5) .. controls +(0,0.8) and +(0,-0.8) .. (1,0.5) .. controls +(0,0.8) and +(0,-0.8) .. (2,1.5);
    \draw[blue,thick] (1,-1.5) node[black,font=\scriptsize,below]{~~$i+1$} .. controls +(0,0.8) and +(0,-0.8) .. (2,-0.5) .. controls +(0,0.8) and +(0,-0.8) .. (2,0.5) .. controls +(0,0.8) and +(0,-0.8) .. (1,1.5);
    \draw[blue,thick] (2,-1.5) node[black,font=\scriptsize,below]{~~~~~~$i+2$} .. controls +(0,0.8) and +(0,-0.8) .. (1,-0.5) .. controls +(0,0.8) and +(0,-0.8) .. (0,0.5) .. controls +(0,0.8) and +(0,-0.8) .. (0,1.5);
    \draw[blue,thick] (3,0) node[blue]{$\cdots$};
  \end{tikzpicture}
  ~~\neq~~
  \begin{tikzpicture}[baseline=-0.3ex,scale=0.5]
    \draw[blue,thick] (-1,0) node[blue]{$\cdots$};
    \draw[blue,thick] (0,-1.5) node[black,font=\scriptsize,below]{$i$} .. controls +(0,0.8) and +(0,-0.8) .. (1,-0.5) .. controls +(0,0.8) and +(0,-0.8) .. (2,0.5) .. controls +(0,0.8) and +(0,-0.8) .. (2,1.5);
    \draw[blue,thick] (1,-1.5) node[black,font=\scriptsize,below]{~~$i+1$} .. controls +(0,0.8) and +(0,-0.8) .. (0,-0.5) .. controls +(0,0.8) and +(0,-0.8) .. (0,0.5) .. controls +(0,0.8) and +(0,-0.8) .. (1,1.5);
    \draw[blue,thick] (2,-1.5) node[black,font=\scriptsize,below]{~~~~~~$i+2$} .. controls +(0,0.8) and +(0,-0.8) .. (2,-0.5) .. controls +(0,0.8) and +(0,-0.8) .. (1,0.5) .. controls +(0,0.8) and +(0,-0.8) .. (0,1.5);
    \draw[blue,thick] (3,0) node[blue]{$\cdots$};
  \end{tikzpicture}
\end{align}
Putting all the above things together, we arrive at the following
abstract definition of the twin group:
\begin{definition*}[Twin group]
  The twin group $T_{n}$ is generated by $n-1$ elements
  $t_{1},\cdots,t_{n-1}$ that satisfy the following relations:
  \begin{subequations}
    \begin{alignat}{3}
      &t_{i}^{2}=1&&\quad\text{for}\quad 1\leq i\leq n-1,&\\
      &t_{i}t_{j}=t_{j}t_{i}&&\quad\text{for}\quad 1\leq i,j\leq n-1\quad\text{and}\quad |i-j|\geq2.&
    \end{alignat}
  \end{subequations}
\end{definition*}

Now, for the application to nonidentical particles, we have to focus
on a special class of planar braids called the \textit{planar pure
  braids}. A planar braid is said to be a planar pure braid if each
strand begins and ends at the same point. A typical example is the
following:
\begin{align}
  (t_{i}t_{i+1})^{3}~~=~~
  \begin{tikzpicture}[baseline=-0.3ex,scale=0.5]
    \draw[blue,thick] (-1,0) node[blue]{$\cdots$};
    \draw[red,thick] (0,-3) node[black,font=\scriptsize,below]{$i$} .. controls +(0,0.8) and +(0,-0.8) .. (0,-2) .. controls +(0,0.8) and +(0,-0.8) .. (1,-1) .. controls +(0,0.8) and +(0,-0.8) .. (2,0) .. controls +(0,0.8) and +(0,-0.8) .. (2,1) .. controls +(0,0.8) and +(0,-0.8) .. (1,2) .. controls +(0,0.8) and +(0,-0.8) .. (0,3);
    \draw[blue,thick] (1,-3) node[black,font=\scriptsize,below]{~$i+1$} .. controls +(0,0.8) and +(0,-0.8) .. (2,-2) .. controls +(0,0.8) and +(0,-0.8) .. (2,-1) .. controls +(0,0.8) and +(0,-0.8) .. (1,0) .. controls +(0,0.8) and +(0,-0.8) .. (0,1) .. controls +(0,0.8) and +(0,-0.8) .. (0,2) .. controls +(0,0.8) and +(0,-0.8) .. (1,3);
    \draw[green!50!black,thick] (2,-3) node[black,font=\scriptsize,below]{~~~~~~$i+2$} .. controls +(0,0.8) and +(0,-0.8) .. (1,-2) .. controls +(0,0.8) and +(0,-0.8) .. (0,-1) .. controls +(0,0.8) and +(0,-0.8) .. (0,0) .. controls +(0,0.8) and +(0,-0.8) .. (1,1) .. controls +(0,0.8) and +(0,-0.8) .. (2,2) .. controls +(0,0.8) and +(0,-0.8) .. (2,3);
    \draw[blue,thick] (3,0) node[blue]{$\cdots$};
  \end{tikzpicture}
\end{align}
Obviously, the set of such planar pure braids forms a subgroup---the
pure twin group $PT_{n}$---because the multiplication of two planar
pure braids is another planar pure braid and the identity element is a
planar pure braid. Mathematically, this subgroup is defined by the
kernel of a surjection from the planar braid group $T_{n}$ on $n$
strands to the symmetric group $S_{n}$ on $n$ letters. Let us next
explain this.

First, by just focusing on the initial and final points, we
immediately see that each planar braid induces a permutation $\sigma$
on $n$ letters $\{1,\cdots,n\}$:
\begin{align}
  \begin{tikzpicture}[baseline=0.65ex,scale=0.5]
    \draw[red,thick] (0,-1) node[black,font=\scriptsize,below]{$1$} .. controls +(0,0.8) and +(0,-0.8) .. (0,0) .. controls +(0,0.8) and +(0,-0.8) .. (1,1) .. controls +(0,0.8) and +(0,-0.8) .. (2,2) node[black,font=\scriptsize,above]{$\sigma(1)$};
    \draw[blue,thick] (1,-1) node[black,font=\scriptsize,below]{$\cdots$} .. controls +(0,0.8) and +(0,-0.8) .. (2,0) .. controls +(0,0.8) and +(0,-0.8) .. (2,1) .. controls +(0,0.8) and +(0,-0.8) .. (1,2) node[black,font=\scriptsize,above]{$\cdots$};
    \draw[green!50!black,thick] (2,-1) node[black,font=\scriptsize,below]{$n$} .. controls +(0,0.8) and +(0,-0.8) .. (1,0) .. controls +(0,0.8) and +(0,-0.8) .. (0,1) .. controls +(0,0.8) and +(0,-0.8) .. (0,2) node[black,font=\scriptsize,above]{$\sigma(n)$};
  \end{tikzpicture}
  ~~\mapsto~~
  \begin{pmatrix}
    1&\cdots&n\\
    \sigma(1)&\cdots&\sigma(n)\\
  \end{pmatrix}
\end{align}
This induction from planar braids to permutations gives a surjection
$f$ from the twin group $T_{n}$ to the symmetric group $S_{n}$ by
$f:t_{i}\mapsto \sigma_{i}$, where $\sigma_{i}=(i,i+1)$ is the
adjacent permutation that interchanges $i$ and $i+1$. Since each
strand begins and ends at the same point, all the planar pure braids
are mapped to the identity element in $S_{n}$ under this surjection
$f$. Hence, $PT_{n}$ is the kernel of $f$. Thus we arrive at the
following definition:
\begin{definition*}[Pure twin group]
  The pure twin group $PT_{n}$ is the kernel of the surjection
  $f:T_{n}\to S_{n}$ given by $t_{i}\mapsto\sigma_{i}$:
  \begin{align}
    PT_{n}=\{t\in T_{n}:f(t)=1\}.
  \end{align}
\end{definition*}

The pure twin group so defined has a special topological meaning. The
key observation is that a tuple of $n$ points $x_{1},\cdots,x_{n}$ in
$\mathbb{R}$ can be identified with a single point
$(x_{1},\cdots,x_{n})$ in $\mathbb{R}^{n}$. In particular, a tuple of
$n$ worldlines in one dimension with no triple-crossings can be
identified with a single trajectory in the $n$-dimensional space with
three-body coincidence points removed. Hence, paths in
$M_{\text{$n$-body}}$ beginning and ending at the same point are in
one-to-one correspondence with planar pure braids on $n$ strands (see
figure \ref{figure:1}). In addition, the multiplication of planar pure
braids coincides with the multiplication in the fundamental group
$\pi_{1}(M_{\text{$n$-body}})$. Thus we arrive at the following:
\begin{proposition*}[Khovanov \cite{Khovanov:1996}]
  The pure twin group $PT_{n}$ is isomorphic to the fundamental group
  of the $n$-body configuration space \eqref{eq:28}:
  \begin{align}
    PT_{n}\cong\pi_{1}(M_{\text{$n$-body}}).
  \end{align}
\end{proposition*}

\printbibliography[heading=bibintoc]
\end{document}